\documentclass[english,prd,superscriptaddress,nofootinbib,preprintnumbers]{revtex4}
\usepackage[latin1]{inputenc}
\usepackage{graphicx}
\usepackage{amssymb}
\usepackage{bm}
\usepackage{color}
\usepackage{amsfonts}
\usepackage{dcolumn}

\def\be{\begin{equation}}
\def\ee{\end{equation}}
\def\ba{\begin{eqnarray}}
\def\ea{\end{eqnarray}}

\begin{document}

\title{Testing for dynamical dark energy models with redshift-space distortions}

\author{Shinji Tsujikawa}
\affiliation{Department of Physics, Faculty of Science, 
Tokyo University of Science, 
1-3, Kagurazaka, Shinjuku-ku, Tokyo 162-8601, Japan}

\author{Antonio De Felice}
\affiliation{TPTP \& NEP, The Institute for Fundamental Study, Naresuan University,
Phitsanulok 65000, Thailand}
\affiliation{Thailand Center of Excellence in Physics, Ministry of Education,
Bangkok 10400, Thailand}

\author{Jailson Alcaniz}
\affiliation{Departamento de Astronomia, Observat\'orio Nacional, 20921-400 Rio de Janeiro - RJ, Brasil}

\date{\today}
\begin{abstract}

The red-shift space distortions in the galaxy power spectrum can be 
used to measure the growth rate of matter density perturbations $\delta_m$.
For dynamical dark energy models in General Relativity
we provide a convenient analytic formula of $f(z) \sigma_8(z)$
written as a function of the redshift $z$, where $f=d \ln \delta_m/d \ln a$
($a$ is the cosmological scale factor) and $\sigma_8$ is the rms amplitude of 
over-density at the scale $8$\,$h^{-1}$\,Mpc.
Our formula can be applied to the models of imperfect fluids, 
quintessence, and k-essence, provided that the dark energy 
equation of state $w$ does not vary significantly and that 
the sound speed is not much smaller than 1.
We also place observational constraints on dark energy models
of constant $w$ and tracking quintessence from 
the recent data of red-shift space distortions.

\end{abstract}
\maketitle

\section{Introduction}

After the first discovery of cosmic acceleration from 
the distance measurements of the supernovae type Ia (SN Ia) \cite{RP}, 
the existence of dark energy has been also supported from other 
observational data such as Cosmic Microwave Background (CMB) \cite{Spergel,Komatsu}
and Baryon Acoustic Oscillations (BAO) \cite{BAO}.
{}From the theoretical point of view, such a late-time cosmic acceleration
is problematic because of the huge difference between 
the observed dark energy scale and the expected value of the vacuum energy 
appearing in particle physics \cite{Weinberg}.
Along with the cosmological constant $\Lambda$, many alternative acceleration 
mechanisms have been proposed, including modifications of the matter/energy content and  
large-scale modifications of gravity
(see Refs.~\cite{review,review2} for reviews).

The dark energy equation of state $w$ is constrained by measuring 
the expansion rate of the Universe from the observations of SN Ia, CMB, 
and BAO \cite{Komatsu,Suzuki,BlakeBAO,BOSS12,DNT12}. 
Although it is possible to rule out some accelerating scenarios from 
the analysis of the cosmic expansion history alone, we require further 
precise observational data to clearly distinguish between models
with subtly-varying $w$. 
So far the $\Lambda$-Cold-Dark-Matter ($\Lambda$CDM) model 
has been consistent with the data, 
but there are many other dynamical models such as 
quintessence \cite{quinpapers}, k-essence \cite{kinf,kespapers}, 
and $f(R)$ gravity \cite{fR} which are compatible 
with current observations.

The large-scale structure of the galaxy distribution provides additional 
important information to discriminate between different dark energy 
models \cite{Tegmark03}. The galaxy clustering occurs due to 
the gravitational instability of primordial matter density perturbations.
The growth rate of matter perturbations can be measured from 
the redshift-space distortion (RSD) of clustering pattern 
of galaxies. This distortion is caused by the peculiar velocity 
of inward collapse motion of the large-scale structure, 
which is directly related to the growth rate of the matter 
density contrast $\delta_m$ \cite{Kaiser}. Hence the RSD measurement 
is very useful to constrain the cosmic growth history.

Recent galaxy redshift surveys have provided bounds on the 
growth rate $f(z)$ or $f(z) \sigma_8 (z)$ in terms of the
redshift $z=1/a-1$, where $f=d \ln \delta_m/d \ln a$
and $\sigma_8$ is the rms amplitude of $\delta_m$
at the comoving scale $8~h^{-1}$ Mpc ($h$ is the normalized 
Hubble constant $H_0=100\,h$\,km\,sec$^{-1}$\,Mpc$^{-1}$) 
\cite{Tegmark04,Percival04,Porto,Guzzo08,Savas,Alam,Blake,Samushia11,Reid12,Beutler12,Koyama, Rapetti,Samushia12}.
Although the observational error bars of $f \sigma_8$ are 
not yet small, the data are consistent with the prediction of the 
$\Lambda$CDM model \cite{Blake,Samushia12}.
Recently the RSD data were used to place constraints
on modified gravity models such as $f(R)$ gravity 
and (extended) Galileons \cite{Linder12,OTT12}. 
Since the growth rate of matter perturbations in these models
is different from that in the $\Lambda$CDM \cite{fRper,Kase,DTextended}, 
the allowed parameter space is quite
tightly constrained even from current observations.

For the models based on General Relativity (GR) without a direct 
coupling between dark energy and non-relativistic matter, the gravitational 
coupling appearing in the matter perturbation equation is 
equivalent to the Newton's gravitational constant, 
as long as the dark energy perturbation is 
negligible relative to the matter perturbation.
Nonetheless the evolution of perturbations depends on the background 
cosmology, so that the dynamical dark energy models with 
$w$ different from $-1$ can be distinguished from 
the $\Lambda$CDM.
In particular, the future RSD observations 
may reach the level of discriminating between different dark energy 
models constructed in the framework of GR.

In this paper we derive an analytic formula of $f(z) \sigma_8(z)$
valid for dynamical dark energy models including 
imperfect fluids, quintessence, and k-essence.
Provided that the sound speed $c_s$ is not much smaller than $1$ 
and that the variation of $w$ is not significant,
our formula can reproduce the full numerical solutions with high accuracy.
The derivation of  $f(z) \sigma_8(z)$ is based on the expansion of $w$
in terms of the dark energy density parameter $\Omega_x$, 
i.e., $w=w_0+\sum_{n=1}w_n (\Omega_x)^n$.
Since $f(z) \sigma_8(z)$ is expressed in terms of the present  
values of $\sigma_8$ and $\Omega_x$ as well as $w_n$ 
($n=0,1,2, \cdots$), our formula is convenient to test
for dynamical dark energy models with the observational data of 
the cosmic growth rate. For the models with constant $w$ there are 3 parameters 
$w_0$, $\sigma_8(z=0)$, and $\Omega_x (z=0)$ in the 
analytic expression of $f(z) \sigma_8(z)$.
In tracking quintessence models \cite{Zlatev}, where 
the dark energy equation of state is nearly constant during the matter 
era ($w=w_{(0)}$), we show that our formula of 
$f(z) \sigma_8(z)$ also contains only 3 parameters: $w_{(0)}$, 
$\sigma_8(z=0)$, and $\Omega_x (z=0)$.
Using the recent RSD data, we carry out the likelihood analysis 
by varying these 3 parameters to find observational bounds 
on $w_0$ and $w_{(0)}$.

This paper is organized as follows.
In Sec.~\ref{cosmodsec} we review cosmological perturbation theory
in general dark energy models including imperfect fluids, quintessence, 
and k-essence. In Sec.~\ref{dynamicalsec} dynamical 
dark energy models are classified depending on how the equation of state 
$w$ is expanded in terms of $\Omega_x$.
In Sec.~\ref{analyticsec} we derive an analytic formula for 
$f(z) \sigma_8(z)$ and in Sec.~\ref{validitysec} 
we confirm the validity of this formula
in concrete examples of dark energy models.
In Sec.~\ref{obserconsec} we perform the likelihood analysis 
to test for constant $w$ and tracking
quintessence models with the recent RSD data. 
Sec.~\ref{consec} is devoted to our main conclusions.

\section{Cosmological perturbations and redshift-space distortions}
\label{cosmodsec}

As the dark energy component we consider a fluid characterized by 
the equation of state $w=P_x/\rho_x$, where $P_x$ is
the pressure and $\rho_x$ is the energy density.
We also take into account non-relativistic matter 
(cold dark matter and baryons) with the 
energy density $\rho_m$ and treat it as a perfect fluid 
with the equation of state $w_m=0$.
We deal with such a two-fluid system in the framework 
of GR under the assumption that 
dark energy is uncoupled to non-relativistic matter.
Our analysis covers quintessence \cite{quinpapers} and 
k-essence \cite{kinf,kespapers} 
models, in which the Lagrangian $P$ of dark energy depends on
a scalar field $\phi$ and a kinetic term $X=-(\nabla \phi)^2/2$.
In these models we have that $P_x=P$ and $\rho_x=2XP_{,X}-P$, 
where $P_{,X} \equiv \partial P/\partial X$.

In the flat Friedmann-Lema\^{i}tre-Robertson-Walker (FLRW) 
background with the scale factor $a(t)$, where $t$ is the cosmic time,
dark energy and non-relativistic matter obey, respectively, the following 
continuity equations 
\ba
& & \rho_x'+3(1+w)\rho_{x}=0\,,\\
& & \rho_m'+3 \rho_m=0\,,
\label{rhophieq}
\ea
where a prime represents a derivative with respect to 
$N=\ln a$.
We introduce the density parameters 
$\Omega_{x}=8 \pi G \rho_{x}/(3H^2)$ and 
$\Omega_{m}=8 \pi G \rho_{m}/(3H^2)$, where
$G$ is the gravitational constant and $H=\dot{a}/a$ 
is the Hubble parameter (a dot represents derivative with respect to $t$).
{}From the Einstein equations it follows that 
\ba
& & \Omega_x+\Omega_m=1\,,
\label{Fri1} \\
& & \frac{H'}{H} = -\frac32 (1+w \Omega_x)\,.
\label{Fri2}
\ea
The dark energy density parameter satisfies
the differential equation 
\be
\Omega_x'=-3w \Omega_{x}(1-\Omega_{x})\,.
\label{Omexeq}
\ee

Let us consider scalar metric perturbations about the flat FLRW background.
We neglect the contribution of tensor and vector perturbations.
In the absence of the anisotropic stress the perturbed line element 
in the longitudinal gauge is given by \cite{Bardeen}
\be
ds^{2}=-(1-2\Phi)\, dt^{2}+a^{2}(t)(1+2\Phi) d {\bm x}^2\,,
\label{permet}
\ee
where $\Phi$ is the gravitational potential.
We decompose the energy densities $\rho_i$ (where $i=x, m$) and 
the pressure $P_x$ into the background and inhomogeneous parts, 
as $\rho_i=\rho_i(t)+\delta \rho_i (t, {\bm x})$ and 
$P_x=P_x(t)+\delta P_x  (t, {\bm x})$. 
We also define the following quantities
\be
\delta_i \equiv \frac{\delta \rho_i}{\rho_i}\,,\qquad
\theta_i \equiv \frac{\nabla^2 v_i}{aH} \qquad
(i=x,m),
\ee
where $v_x$ and $v_m$ are the rotational-free velocity potentials 
of dark energy and non-relativistic matter, respectively.

In Fourier space dark energy perturbations obey 
the following equations of motion \cite{Kodama,Bean,Sapone,DEbook}
\ba
& & \delta_x'+3(c_x^2-w)\delta_x
=-(1+w) \left( 3\Phi'+\theta_x \right)\,,
\label{de1}\\
& & \theta_x'+\left( 2-3w+\frac{H'}{H}+
\frac{w'}{1+w} \right) \theta_x=\left( \frac{k}{aH} \right)^2
\left( \frac{c_x^2}{1+w} \delta_x-\Phi \right)\,,
\label{de2}
\ea
where $c_x^2=\delta P_{x}/\delta \rho_x$ and $k$
is a comoving wave number.
The perturbed equations of non-relativistic 
matter (perfect fluids) are
\ba
& & \delta_m'=-3 \Phi'-\theta_m\,,
\label{dm1}\\
& & \theta_m'+\left( 2+\frac{H'}{H} \right)\theta_m=
-\left( \frac{k}{aH} \right)^2 \Phi\,.
\label{dm2}
\ea
{}From the Einstein equations we obtain
\ba
& & \left( \frac{k}{aH} \right)^2 \Phi=
\frac32 \left( \Omega_m \hat{\delta}_m+
\Omega_x \hat{\delta}_x \right)\,,
\label{Ein1} \\
& & \Phi' + \Phi=-\frac32 \left( \frac{aH}{k} \right)^2
\left[ \Omega_m \theta_m+(1+w) \Omega_x \theta_x \right]\,,
\label{Ein2}
\ea
where $\hat{\delta}_m$ and $\hat{\delta}_x$ are the rest frame 
gauge-invariant density perturbations defined by 
\be
\hat{\delta}_m=\delta_m+3\left( \frac{aH}{k} \right)^2 
\theta_m\,,\qquad
\hat{\delta}_x=\delta_x+3\left( \frac{aH}{k} \right)^2 
(1+w) \theta_x\,.
\label{hatqu}
\ee

For imperfect fluids such as quintessence and k-essence
there exist non-adiabatic entropy perturbations generated 
from dissipative processes.
We introduce a gauge-invariant entropy perturbation $\delta s_x$
of dark energy, as \cite{Kodama,Bean,BTW05}
\be
\delta s_x=(c_x^2-c_a^2) \delta_x=
\frac{\dot{P}_x}{\rho_x} \left( \frac{\delta P_x}
{\dot{P}_x}-\frac{\delta \rho_x}{\dot{\rho}_x} \right)\,,
\ee
where $c_a$ is the adiabatic sound speed defined by 
\be
c_a^2=\frac{\dot{P}_x}{\dot{\rho}_x}
=w-\frac{w'}{3(1+w)}\,.
\ee
In the rest frame where the entropy perturbation is 
given by $\delta s_x=(\hat{c}_x^2-c_a^2)\hat{\delta}_x$, 
the sound speed squared $c_s^2 \equiv \hat{c}_x^2$ is 
gauge-invariant. Using the definition of $\hat{\delta}_x$
in Eq.~(\ref{hatqu}), the pressure perturbation of dark
energy can be expressed as
\be
\delta P_x=c_s^2\,\delta \rho_x+3\left( \frac{aH}{k} \right)^2 
(1+w) (c_s^2-c_a^2) \rho_x \theta_x\, ,
\label{delP}
\ee
whereas the sound speed squared $c_x^2=\delta P_x/\delta \rho_x$ 
in the random frame is related to $c_s^2$ via
\be
c_x^2=c_s^2+3\left( \frac{aH}{k} \right)^2
(1+w) (c_s^2-c_a^2) \frac{\theta_x}{\delta_x}\,.
\label{delP2}
\ee
In terms of $c_s^2$, Eqs.~(\ref{de1}) and (\ref{de2}) 
can be written as
\ba
& & \delta_x'+3(c_s^2-w)\delta_x
=-3(1+w) \Phi'-(1+w) 
\left[ 1+9 \left( \frac{aH}{k} \right)^2
(c_s^2-c_a^2) \right] \theta_x\,,
\label{de3}\\
& & \theta_x'+\left( 2+\frac{H'}{H}-3c_s^2
\right) \theta_x=\left( \frac{k}{aH} \right)^2
\left( \frac{c_s^2}{1+w} \delta_x-\Phi \right)\,.
\label{de4}
\ea

In k-essence characterized by the Lagrangian density $P(\phi,X)$,
the perturbed quantities can be expressed as
\ba
\hspace{-0.5cm} 
\delta \rho_{x} &=&
\left( P_{,X}+2XP_{,XX} \right) \delta X-
\left( P_{,\phi}-2X P_{,X \phi} \right) \delta \phi\,,
\label{delrho}\\
\hspace{-0.5cm} 
\delta P_{x} &=& P_{,X} \delta X+P_{,\phi} \delta \phi\,,
\label{delP3}\\
\hspace{-0.5cm} 
\theta_{x} &=& \frac{k^2}{a \dot{\phi}} \delta \phi\,,
\ea
where $\delta \phi$ is the field perturbation and 
$\delta X=\dot{\phi}\,\dot{\delta \phi}+\dot{\phi}^2 \Phi$.
{}From Eq.~(\ref{delP}) the rest frame sound speed
can be obtained by setting $\delta \phi=0$ in 
Eqs.~(\ref{delrho}) and (\ref{delP3}), i.e.,  
$c_s^2=P_{,X}/(P_{,X}+2XP_{,XX})$ \cite{Garriga}.
In quintessence characterized by the Lagrangian 
$P=X-V(\phi)$, the sound speed squared 
reduces to $c_s^2=1$.

From Eqs.~(\ref{dm1}) and (\ref{dm2})
it follows that
\begin{equation}
\delta_m''+\left( 2+\frac{H'}{H} \right) \delta_m'
-\left( \frac{k}{aH} \right)^2 \Phi
=-3 \left[ \Phi''+\left( 2+\frac{H'}{H} \right)\Phi' \right]\,.
\label{delmeq0}
\end{equation}
For the perturbations deep inside the Hubble radius ($k \gg aH$)
relevant to large-scale structures, the r.h.s. of Eq.~(\ref{delmeq0}) 
can be neglected relative to the l.h.s. of it, in addition to the 
fact that $\hat{\delta}_m \simeq \delta_m$.
If the contribution of dark energy perturbations can be neglected
relative to that of matter perturbations in Eq.~(\ref{Ein1}), i.e. 
$|\Omega_m \hat{\delta}_m| \gg |\Omega_x \hat{\delta}_x|$,
Eq.~(\ref{delmeq0}) reads
\be
\delta_m''+\frac12 \left(1-3w \Omega_x \right) \delta_m'
-\frac32 \Omega_m \delta_m \simeq 0\,,
\label{delmeq}
\ee
where we made use of Eq.~(\ref{Fri2}).

During the deep matter era in which $\Omega_m \simeq 1$, 
there is a growing-mode solution 
$\delta_m=\delta_m' \propto a$ to Eq.~(\ref{delmeq}).
In this regime Eq.~(\ref{Ein1}) tells us that $\Phi=\,{\rm constant}$ and 
hence $\theta_m \simeq -\delta_m'$ from Eq.~(\ref{dm1}).
For the dark energy density contrast $\delta_x$, it is natural to 
choose the adiabatic initial condition \cite{Kodama,DEbook}
\be
\delta_x=(1+w) \delta_m\,.
\label{adi}
\ee
The initial condition of $\theta_x$ is known by substituting 
Eq.~(\ref{adi}) and $\delta_x'=(1+w+w')\delta_m$ into 
Eq.~(\ref{de3}).
We will discuss the accuracy of the approximate 
equation (\ref{delmeq}) in Sec.~\ref{dynamicalsec}.

\begin{table}[t]
\begin{center}
\begin{tabular}{|c|c|c|c|}
\hline
$z$ &  $f\sigma_8$ & survey \\
\hline
\hline
0.067 & 0.423$\pm$ 0.055 & 6dFGRS (2012) \cite{Beutler12} \\
\hline
0.17 & 0.51$\pm$ 0.06 & 2dFGRS (2004) \cite{Percival04} \\
\hline
0.22 & 0.42$\pm$ 0.07 & WiggleZ (2011) \cite{Blake} \\
\hline
0.25 & 0.3512$\pm$ 0.0583 & SDSS LRG (2011) \cite{Samushia11} \\
\hline
0.37 & 0.4602$\pm$ 0.0378 & SDSS LRG (2011) \cite{Samushia11} \\
\hline
0.41 & 0.45$\pm$ 0.04 & WiggleZ (2011) \cite{Blake} \\
\hline
0.57 & 0.415$\pm$ 0.034 & BOSS CMASS (2012) \cite{Reid12} \\
\hline
0.6 & 0.43$\pm$ 0.04 & WiggleZ (2011) \cite{Blake} \\
\hline
0.78 & 0.38$\pm$ 0.04 & WiggleZ (2011)\cite{Blake} \\
\hline
\end{tabular}
\end{center}
\caption[fs8]{Data of $f\sigma_8$ versus the redshift $z$
measured from the RSD.}
\label{table:fs8constraints}
\end{table}

The growth rate of matter perturbations can be measured from 
the RSD in clustering pattern of galaxies because of radial
peculiar velocities.
The perturbation $\delta_g$ of galaxies is related 
to the matter perturbation $\delta_m$, as
$\delta_g=b \delta_m$, where $b$ is a bias factor.
The galaxy power spectrum ${\cal P}_g ({\bm k})$ in 
the redshift space can be expressed 
as \cite{Kaiser,Tegmark06,Song09,White}
\begin{equation}
{\cal P}_g ({\bm k})={\cal P}_{gg} ({\bm k})
-2\mu^2 {\cal P}_{g \theta} ({\bm k})
+\mu^4 {\cal P}_{\theta \theta} ({\bm k})\,,
\label{Pred}
\end{equation}
where $\mu={\bm k} \cdot {\bm r}/(kr)$ is the cosine 
of the angle of the ${\bm k}$ vector to the line 
of sight (vector ${\bm r}$),
${\cal P}_{gg} ({\bm k})$ and ${\cal P}_{\theta \theta} ({\bm k})$ 
are the real space power spectra 
of galaxies and $\theta$, respectively, 
and ${\cal P}_{g \theta} ({\bm k})$ is the cross 
power spectrum of galaxy-$\theta$ fluctuations
in real space.

In Eq.~(\ref{dm1}) the variation of the gravitational potential is 
neglected relative to the growth rate of $\delta_m$, so that 
\begin{equation}
\theta_m \simeq -f \delta_m\,,\qquad
f \equiv \frac{\dot{\delta}_m}{H \delta_m}\,.
\label{continuity}
\end{equation}
Under the continuity equation (\ref{continuity}), 
${\cal P}_{gg}$, ${\cal P}_{g \theta}$, and ${\cal P}_{\theta \theta}$ 
in Eq.~(\ref{Pred}) depend on $(b \delta_m)^2$, 
$(b \delta_m)(f \delta_m)$, and $(f \delta_m)^2$, respectively.
We normalize the amplitude of $\delta_m$ at the 
scale $8\,h^{-1}$~Mpc, for which we write $\sigma_8$.
Assuming that the growth of perturbations is scale-independent, 
the constraints on $b \delta_m$ and $f \delta_m$ translate into 
those on $b \sigma_8$ and $f \sigma_8$.
The advantage of using $f \sigma_8$ is that the growth rate
is directly known without the bias factor $b$.
In Table \ref{table:fs8constraints} we show the current data of $f\sigma_8$ as a function of $z$ from the RSD measurements.

\section{Dynamical dark energy models}
\label{dynamicalsec}

In this section, we discuss a number of dynamical dark energy models in which
the field equations presented in the previous section can be applied.

\subsection{Imperfect fluids}
\label{impersec}

For imperfect fluids the rest frame sound speed $c_s$
is generally different from the adiabatic sound speed $c_a$.
For constant $w$ one has $c_a^2=w$.
If $c_s^2$ is constant as well, the evolution of dark energy 
perturbations is known by solving Eqs.~(\ref{de3}) and (\ref{de4})
together with the background equations (\ref{Fri1})-(\ref{Omexeq}).
This is the approach taken in Ref.~\cite{Bean}.
Note also that for $c_s$ of the order of 1 the contribution of dark energy 
perturbations to $\Phi$ in Eq.~(\ref{Ein1}) is negligibly small
relative to matter perturbations \cite{Bean,Riotto}.

\begin{figure}
\includegraphics[height=3.2in,width=3.4in]{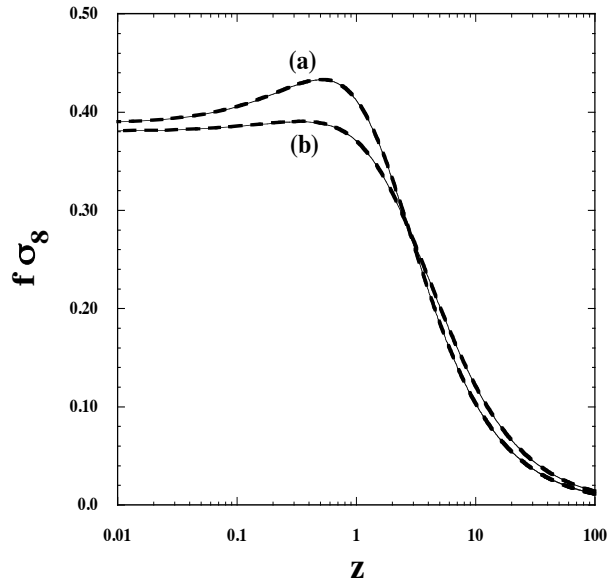}
\caption{\label{dperfig}
Evolution of $f\sigma_8$ versus the redshift $z$ for $c_s^2=1$,
$k=0.07\,h$~Mpc$^{-1}$, $\sigma_8 (z=0)=0.811$, and 
$\Omega_m (z=0)=0.73$.
The cases (a) and (b) correspond to $w=-0.8$ and $w=-0.5$,
respectively. The solid lines show the numerically integrated solutions, 
whereas the bald dashed lines correspond to the solutions derived
by using the approximate equation (\ref{delmeq}).}
\end{figure}

In Fig.~\ref{dperfig} we plot the evolution of $f \sigma_8$
for $c_s^2=1$ and $k=0.07\,h$~Mpc$^{-1}$
with two different values of $w$.
The approximate equation (\ref{delmeq}) reproduces 
the full numerical solution within the 0.1 \% accuracy. 
This means that, for $c_s^2=1$, the contribution of dark energy 
perturbations to the gravitational potential 
is suppressed relative to that of matter perturbations.
For $c_s^2=0$ and $w=-0.8$ we find that the difference of 
$f \sigma_8(z=0)$ between
the numerical result and the approximated solution is about 4\,\%
for the mode $k=0.07\,h$~Mpc$^{-1}$.
For $w$ larger than $-0.8$ the difference gets even larger, 
but such values of $w$ are not allowed observationally \cite{Komatsu,Suzuki}. 
In Ref.~\cite{Sapone} it was found that galaxy redshift and 
tomographic redshift surveys can constrain the sound speed
only if $c_s^2$ is sufficiently small, of the order 
of $c_s^2 \lesssim 10^{-4}$.

\subsection{Quintessence}

Quintessence \cite{quinpapers} is characterized by the Lagrangian 
$P(\phi, X)=X-V(\phi)$, where $V(\phi)$ is the field potential. 
In this case the sound speed $c_s$ is equivalent to 1, 
so that the contribution of dark energy perturbations 
to the gravitational potential is negligible.
There are several classes of potentials which 
give rise to different evolution of $w$.

The first class is the model with constant $w$, 
which can be realized by the following 
potential \cite{review,Saini03,Sahni06}
\be
V(\phi) = \frac{3H_0^2 (1-w) (1-\Omega_{m0})^{1/|w|}}
{16 \pi G \Omega_{m0}^\alpha} \sinh^{-2\alpha} \left( |w| \sqrt{\frac{6\pi G}
{1+w}} (\phi-\phi_0+\phi_1) \right),
\label{quinpo1}
\ee
where $\alpha=(1+w)/|w|$, $\phi_0$ and $\Omega_{m0}$ are the 
today's values of $\phi$ and $\Omega_m$ respectively, and 
\be
\phi_1=\sqrt{\frac{1+w}{6\pi G}} \frac{1}{|w|}
\ln \frac{1+\sqrt{1-\Omega_{m0}}}{\sqrt{\Omega_{m0}}}\,.
\ee
This case is identified as an imperfect fluid 
with $c_s^2=1$ and $w=\,$constant 
discussed in Sec.~\ref{impersec}.

The second class consists of freezing models \cite{CL05}, 
in which the evolution of the field gradually slows down
at late times. A typical example is the inverse 
power-law potential \cite{Ratra}
\be
V(\phi)=M^{4+p} \phi^{-p}\,,
\label{inversepo}
\ee
where $p$ is a positive constant.
For this potential there exists a so-called tracker 
solution \cite{Zlatev} with a nearly constant field equation of state
$w=w_{(0)} \equiv -2/(p+2)$ during the matter era, 
which is followed by the decrease of $w$.
Considering a homogeneous perturbation $\delta w$ around 
$w_{(0)}$, the field equation of state is expressed as \cite{Chiba}
\be
w = w_{(0)}+ \sum_{n=1}^{\infty} 
\frac{(-1)^{n-1} w_{(0)} (1-w_{(0)}^2)}
{1-(n+1)w_{(0)}+2n(n+1) w_{(0)}^2}
\left( \frac{\Omega_{x}}{1-\Omega_{x}}
\right)^{n}\,. 
\label{wtrack}
\ee
Expansion of $w$ around $\Omega_{x}$ reads
\ba
w &=& w_{(0)}+\frac{(1-w_{(0)}^2)w_{(0)}}{1-2w_{(0)}+4w_{(0)}^2}
\Omega_{x}+\frac{(1-w_{(0)}^2)w_{(0)}^2 (8w_{(0)}-1)}
{(1-2w_{(0)}+4w_{(0)}^2)(1-3w_{(0)}+12w_{(0)}^2)} 
\Omega_{x}^2 \nonumber \\
& &+\frac{2(1-w_{(0)}^2)w_{(0)}^3 (4w_{(0)}-1)(18w_{(0)}+1)}
{(1-2w_{(0)}+4w_{(0)}^2)(1-3w_{(0)}+12w_{(0)}^2)
(1-4w_{(0)}+24w_{(0)}^2)} \Omega_{x}^3
+{\cal O} (\Omega_{x}^4)\,,
\label{wtrack2}
\ea
which varies with the growth of $\Omega_{x}$.

The third class consists of thawing models \cite{CL05}, 
in which the field is nearly frozen by a Hubble friction during the early
cosmological epoch and it starts to evolve once
the field mass $m$ drops below $H$. In this case $w$ 
is initially close to $-1$ and then $w$ starts to grow at late times. 
The representative potential of this class is that 
of the pseudo Nambu-Goldstone boson \cite{PNGB}, i.e., 
\be
V(\phi)=\Lambda^4 \left[ 1+\cos (\phi/f) \right]\,,
\ee
where $\Lambda$ and $f$ are constants.
Assuming that the evolution of the scale factor can be 
approximated as that of the $\Lambda$CDM model, 
the field equation of state is estimated 
as \cite{Dutta} (see also Ref.~\cite{Scherrer})
\be
w =-1+(1+\tilde{w}_0) a^{3(K-1)} \left[
\frac{(K-F(a))(F(a)+1)^K+(K+F(a))(F(a)-1)^K}
{(K-\Omega_{x 0}^{-1/2})(\Omega_{x 0}^{-1/2}+1)^K
+(K+\Omega_{x 0}^{-1/2})(\Omega_{x 0}^{-1/2}-1)^K}
\right]^2\,,
\label{wtha}
\ee
where $\tilde{w}_0$ and $\Omega_{x 0}$ are the today's values 
of $w$ and $\Omega_{x}$ respectively, and 
\be
K=\sqrt{1-\frac{V_{,\phi \phi} (\phi_i)}
{6 \pi G V(\phi_i)}}\,,\qquad
F(a) = \sqrt{1+(\Omega_{x 0}^{-1}-1)a^{-3}}\,.
\ee
The constant $K$ is related to the mass of the field
at the initial displacement, $\phi=\phi_i$.
Expansion around $\Omega_{x}=0$ gives 
\be
w=-1+w_1 \Omega_{x}
+\frac15 (K^2+1)w_1 \Omega_x^2
+\frac{1}{175} (3K^4+31K^2+3)w_1 \Omega_x^3
+{\cal O}(\Omega_x^4)\,,
\label{wthaw2}
\ee
where 
\be
w_1=\frac{4}{9}\, \frac{1+\tilde{w}_0}
{[(K-\Omega_{x 0}^{-1/2})(\Omega_{x 0}^{-1/2}+1)^K
+(K+\Omega_{x 0}^{-1/2})(\Omega_{x 0}^{-1/2}-1)^K]^2} 
\left( \frac{1-\Omega_{x 0}}
{\Omega_{x 0}} \right)^{K-1} 
K^2 (K-1)^2 (K+1)^2\,.
\ee
The growth of $\Omega_{x}$ leads to the 
deviation from $w=-1$.

\subsection{K-essence}

The equation of state for K-essence with the Lagrangian density 
$P(\phi, X)$ is given by $w=P/(2XP_{,X}-P)$.
This shows that cosmic acceleration with $w \approx -1$ 
can be realized either for (a) $X \approx 0$ or (b) $P_{,X} \approx 0$.

In the case (a) Chiba {\it et al.} \cite{Chibake} showed that, for the 
factorized function $P(\phi,X)=V(\phi)F(X)$, the field equation of 
state is given by the same form as Eq.~(\ref{wtha})
with the replacement $K=\sqrt{1-V_{,\phi \phi}(\phi_i)/
(6\pi G F_{,X}(0) V^2(\phi_i))}$, where $F(X)$ is expanded 
around $X=0$ as $F(X)=F(0)+F_{,X}(0)X$.
In this case the sound speed squared is also close to 1, 
so that the situation is similar to that in thawing quintessence models.

In the case (b) the evolution of $w$ depends on the functional form 
of $P(\phi,X)$, so it is difficult to derive a general expression 
of $w$ \cite{Chibake}.
One of the examples which belongs to this class is the dilatonic 
ghost condensate model \cite{Piazza} characterized by the Lagrangian
$P(\phi, X)=-X+e^{\kappa \lambda \phi}X^2/M^4$, where 
$\kappa=\sqrt{8\pi G}$, $\lambda$ and $M$ are 
constants (see also Ref.~\cite{Mukohyama}).
In this model the fixed points during the radiation and matter 
eras correspond to $P_{,X}=0$ and $w=-1$, i.e.,
$y \equiv Xe^{\kappa \lambda \phi}/M^4=1/2$.
On the other hand the (no-ghost) accelerated fixed point corresponds to 
$y=1/2+\lambda^2 f(\lambda)/16$ with the equation of state 
$w=-[1-\lambda^2 f(\lambda)/8]/[1+3\lambda^2 f(\lambda)/8]>-1$, 
where $f(\lambda)=1+\sqrt{1+16/(3\lambda^2)}$.
Hence the evolution of $w$ is similar to that 
in thawing quintessence models.

The sound speed squared in the dilatonic ghost condensate model
is given by $c_s^2=(2y-1)/(6y-1)$, so that $c_s^2 \simeq 0$
during radiation and matter eras.
The late-time cosmic acceleration occurs for $1/2 \le y<2/3$
and hence $0 \le c_s^2<1/9$ at this fixed point.
The fact that $c_s^2$ is close to 0 during most of the cosmological 
epoch is a different signature relative to quintessence.
However, since $w$ is very close to $-1$ during radiation and matter 
eras, the adiabatic initial condition (\ref{adi}) shows that  
the dark energy perturbation $\delta_x$ is initially 
suppressed relative to the matter perturbation $\delta_m$.
As long as the today's value of $w$ is not significantly 
away from $-1$ the contribution of the dark energy perturbation 
to $\Phi$ is suppressed relative to the matter perturbation, 
so that the approximate equation (\ref{delmeq}) 
can be trustable even in such cases.

\section{Analytic solutions of $f \sigma_8$}
\label{analyticsec}

We derive analytic solutions of $f \sigma_8$ by solving 
the approximate equation (\ref{delmeq}).
Recall that the equation of state for tracking and thawing 
models of quintessence can be expressed in terms of
the field density parameter $\Omega_x$, see 
Eqs.~(\ref{wtrack2}) and (\ref{wthaw2}).
We generally expand the dark energy equation of state in terms of
the density parameter $\Omega_{x}$, as
\be
w=w_0+\sum_{n=1}^{\infty} w_n (\Omega_{x})^n\,.
\label{wexpand}
\ee
Since $\Omega_{x}$ grows as large as 0.7 today, we expect 
that it may be necessary to pick up the terms higher than 
the first few terms in Eq.~(\ref{wexpand}).

In terms of the function $f=\delta_m'/\delta_m$, 
Eq.~(\ref{delmeq}) can be written as \cite{Wang}
\begin{equation}
3 w \Omega_{x} (1-\Omega_{x}) \frac{df}{d\Omega_{x}}
=f^2+\frac12 (1-3w \Omega_{x})f-\frac32 (1-\Omega_{x})\,,
\label{fOme}
\end{equation}
where we employed Eq.~(\ref{Omexeq}).
Introducing the growth index $\gamma$ as $f=(\Omega_m)^{\gamma}
=(1-\Omega_{x})^{\gamma}$ \cite{Peebles}, Eq.~(\ref{fOme}) reads
\begin{equation}
3w \Omega_{x} (1-\Omega_{x}) 
\ln  (1-\Omega_{x})
\frac{d \gamma}{d \Omega_{x}} 
=\frac12-\frac32 w(1-2\gamma)\Omega_{x}
+(1-\Omega_{x})^{\gamma}-\frac32 
(1-\Omega_{x})^{1-\gamma}\,.
\label{gamOme}
\end{equation}

We derive the solution of Eq.~(\ref{gamOme}) 
by expanding $\gamma$ in terms of $\Omega_{x}$, i.e.,
$\gamma=\gamma_0+\sum_{n=1} \gamma_n (\Omega_{x})^n$.
While $\Omega_m$ is smaller than $\Omega_x$ today, the 
former is not suitable as an expansion parameter as we 
would like to derive an analytic formula valid at high 
redshifts as well. In fact, it is expected that future RSD surveys
such as Subaru/FMOS will provide high-redshift data up to $z=2$.
Using the expansion of $w$ in Eq.~(\ref{wexpand}) as well,
we obtain the following approximate solution 
\begin{eqnarray}
\gamma &=& \frac{3(1-w_0)}{5-6w_0}+\frac32 
\frac{(1-w_0)(2-3w_0)+2w_1(5-6w_0)}{(5-6w_0)^2 (5-12w_0)}
\Omega_{x} \nonumber \\
&&+[(w_0-1)(3w_0-2)(324w_0^2-420w_0+97)+12(5-6w_0)^2
(5w_2-12w_2w_0+12w_1^2) \nonumber \\
& &~~~+6(5-6w_0)(72w_0^2-90w_0+23)w_1]/
[4(5-6w_0)^3 (5-12w_0)(5-18w_0)]\,\Omega_{x}^2
+{\cal O} (\Omega_{x}^3)\,.
\label{gammaap}
\end{eqnarray}
The 1-st order solution is identical to the one found 
in Ref.~\cite{Gong} by setting $w_1=0$.
For $w_0=-1$, $w_1=0$, and $w_2=0$ it follows that 
$\gamma \simeq 0.545+7.29 \times 10^{-3} \Omega_{x}+
4.04 \times 10^{-3}\Omega_{x}^2$.
In this case the second and third terms are indeed much smaller than the 
first one, so that $\gamma$ is nearly constant.
For the models with $w_0=-1$, $w_1=0.3$, and $w_2=0$ 
(in which case the value of $w$ today is around $-0.8$)
one has $\gamma \simeq 0.545+1.21 \times 10^{-2} \Omega_{x}
+6.55 \times 10^{-3} \Omega_{x}^2$.
Even in this case the variation of $\gamma$ induced by the second 
and third terms is small (see also Refs.~\cite{Linder} for 
related works).

{}From the definition of $f$ the matter perturbation obeys
the differential equation $(\ln \delta_m)'=(1-\Omega_{x})^{\gamma}$.
Using Eq.~(\ref{Omexeq}), we obtain
\begin{equation}
\frac{d}{d \Omega_{x}} \ln \delta_m
=-\frac{(1-\Omega_{x})^{\gamma-1}}{3w \Omega_{x}}\,.
\label{delmeqa}
\end{equation}
In the following we derive the solution of this equation under 
the approximation that $\gamma$ is constant.
We expand the term $(1-\Omega_{x})^{\gamma-1}$ around 
$\Omega_{x}=0$, as 
\begin{equation}
(1-\Omega_{x})^{\gamma-1}=1+\sum_{n=1}^{\infty} \alpha_n
(\Omega_{x})^n\,,\qquad 
\alpha_n=\frac{(-1)^n}{n!}
(\gamma-1)(\gamma-2) \cdots (\gamma-n)\,.
\end{equation}
In order to evaluate the r.h.s. of Eq.~(\ref{delmeqa}), 
we expand $1/w$ in the form
\begin{equation}
\frac{1}{w}=\frac{1}{w_0} \left[ 1+\sum_{n=1}^{\infty} 
\beta_{n} (\Omega_{x})^n \right]\,,
\end{equation}
where the coefficients $\beta_n$'s can be expressed by $w_n$'s, 
say $\beta_1=-w_1/w_0$.
Then Eq.~(\ref{delmeqa}) can be written as
\begin{equation}
\frac{d}{d \Omega_{x}} \ln \delta_m
=-\frac{1}{3w_0 \Omega_{x}}
\left[ 1+\sum_{n=1}^{\infty} c_n(\Omega_{x})^n \right]\,,
\label{delmeqa2}
\end{equation}
where
\begin{equation}
c_n=\sum_{i=0}^n \alpha_{n-i} \beta_{i}\,,
\label{cndef}
\end{equation}
with $\alpha_0=\beta_0=1$.
The first three coefficients $c_i$ are 
\begin{eqnarray}
c_1 &=& -(\gamma-1)-\frac{w_1}{w_0}\,,\label{c1} \\
c_2 &=& \frac12 (\gamma-1)(\gamma-2)+
(\gamma-1) \frac{w_1}{w_0}-\frac{w_2 w_0-w_1^2}{w_0^2}\,,\\
c_3 &=& -\frac16 (\gamma-1)(\gamma-2)(\gamma-3)
-\frac12 (\gamma-1)(\gamma-2) \frac{w_1}{w_0}
+(\gamma-1)\frac{w_2 w_0-w_1^2}{w_0^2}
-\frac{w_3 w_0^2-2w_1 w_2 w_0+w_1^3}{w_0^3}\,.
\label{c3} 
\end{eqnarray}

Integrating Eq.~(\ref{delmeqa2}), it follows that 
\begin{equation}
\delta_m=\delta_{m0} \exp \left\{ \frac{1}{3w_0}
\left[ \ln \frac{\Omega_{x 0}}{\Omega_{x}}+\sum_{n=1}^{\infty}
\frac{c_n}{n} \left( (\Omega_{x 0})^n- (\Omega_{x})^n \right) \right]
\right\},
\end{equation}
where $\delta_{m0}$ is the today's value of $\delta_m$. 
Normalizing $\delta_{m0}$ in terms of $\sigma_8 (z=0)$, 
we obtain the following expression 
\begin{equation}
f \sigma_8 (z)=
(1-\Omega_{x})^{\gamma}\,\sigma_8 (z=0)\, \exp \left\{ \frac{1}{3w_0}
\left[ \ln \frac{\Omega_{x 0}}{\Omega_{x}}+\sum_{n=1}^{\infty}
\frac{c_n}{n} \left( (\Omega_{x 0})^n- (\Omega_{x})^n \right) \right]
\right\}\,.
\label{fsigma8}
\end{equation}
In terms of the redshift $z$ the energy densities of 
non-relativistic matter and dark energy
are given, respectively, by $\rho_m=\rho_{m0} (1+z)^3$ and 
$\rho_{x}=\rho_{x 0} \exp [\int_{0}^z 3(1+w)/(1+\tilde{z}) d \tilde{z}]$.
The 0-th order solution to the field energy density is obtained 
by substituting $w=w_0$ into the expression of $\rho_{x}$, i.e., 
$\rho_{x}^{(0)}=\rho_{x 0}(1+z)^{3(1+w_0)}$.
This gives the 0-th order solution to $\Omega_{x}$, as
$\Omega_{x}^{(0)}=\Omega_{x 0} (1+z)^{3w_0}/
[1-\Omega_{x 0}+\Omega_{x 0} (1+z)^{3w_0}]$.
If we expand $w$ up to first order with respect to $\Omega_{x}$, 
we can use the iterative solution 
$w=w_0+w_1 \Omega_{x}^{(0)}$. 
This process leads to the following integrated solution 
of $\Omega_{x}$ :
\begin{equation}
\Omega_{x}^{(1)}=\frac{\Omega_{x 0} (1+z)^{3w_0}
[1-\Omega_{x 0}+\Omega_{x 0} (1+z)^{3w_0}]^{w_1/w_0}}
{1-\Omega_{x 0}+\Omega_{x 0} (1+z)^{3w_0}
[1-\Omega_{x 0}+\Omega_{x 0} (1+z)^{3w_0}]^{w_1/w_0}}\,.
\label{Omegax1}
\end{equation}
In the presence of the terms higher than second order, 
we can simply carry out the similar iterative processes. 
Practically it is sufficient to use the 1-st order solution 
(\ref{Omegax1}) for the evaluation of $\Omega_x$ 
in Eq.~(\ref{fsigma8}).

The growth factor $\gamma$ in Eq.~(\ref{fsigma8}) is given 
by the analytic formula (\ref{gammaap}). 
Since $\gamma$ is expressed in terms of 
$w_n$ ($n=0,1,\cdots$) and $\Omega_x$, this means that 
$f\sigma_8(z)$ depends on the free parameters 
$w_i$, $\Omega_{x0}$, and $\sigma_8 (z=0)$. 
For the models of constant equation 
of state there are 3 free parameters $w_0$, $\Omega_{x0}$, 
$\sigma_8 (z=0)$ in the expression of $f\sigma_8(z)$.

In tracking quintessence models the coefficients 
$w_n$'s ($n \geq 1$) are expressed in terms of 
$w_0=w_{(0)}$, see Eq.~(\ref{wtrack2}).
Hence there are only 3 free parameters $w_0$, 
$\Omega_{x0}$, and $\sigma_8(z=0)$. 
In thawing models of quintessence one has 
$w_0=-1$ and $w_n$'s ($n \geq 2$) can be expressed 
in terms of $w_1$ and $K$, see Eqs.~(\ref{wthaw2}). 
Then $f\sigma_8$ in Eq.~(\ref{fsigma8}) is written as a function 
of $z$ with 4 free parameters: 
$w_1$, $\Omega_{x0}$, $\sigma_8(z=0)$, and $K$.

\section{Validity of analytic solutions}
\label{validitysec}

We study the validity of the analytic estimation given in 
the previous section.
We discuss three different cases: (i) constant $w$ models, 
(ii) tracking models, and (iii) thawing models, separately.
In all the numerical simulations in this section, we identify 
the present epoch to be $\Omega_{x0}=0.73$ with
$\sigma_8 (z=0)=0.811$.

\subsection{Constant $w$ models}
\label{conwsec}
\begin{figure}
\includegraphics[height=3.1in,width=3.3in]{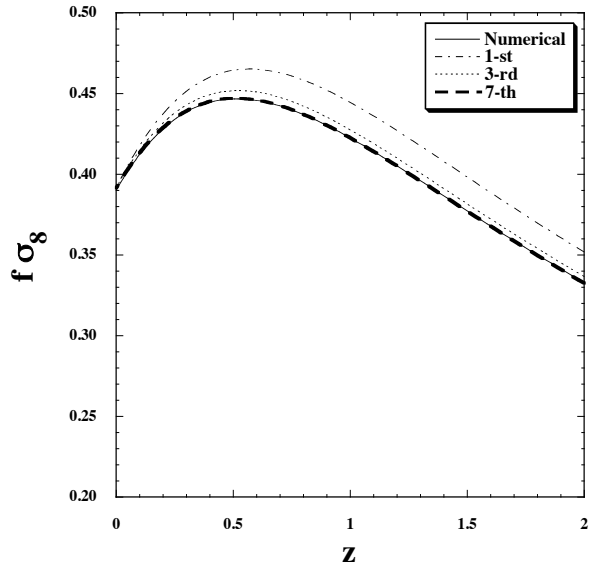}
\caption{\label{worder}
Evolution of $f\sigma_8$ versus $z$ for the constant $w$ model 
with $w=-0.9$ and $c_s^2=1$.
The present epoch is identified to be $\Omega_{x 0}=0.73$
with $\sigma_8 (z=0)=0.811$.
The solid line shows the numerically integrated solution, 
whereas the dot-dashed, dotted, and bold dashed lines
correspond to the solutions summed up to 1-st, 3-rd, and 
7-th order terms of $c_n$ in Eq.~(\ref{fsigma8}).}
\end{figure}
\begin{figure}
\includegraphics[height=3.2in,width=3.4in]{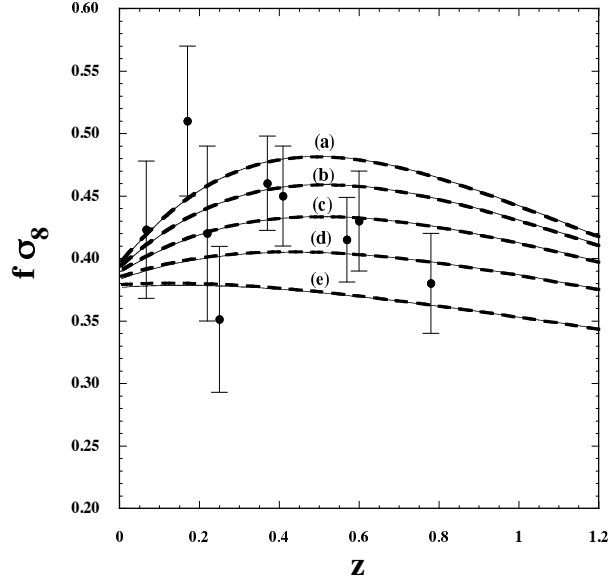}
\caption{\label{fsigmafig}
Evolution of $f\sigma_8$ versus $z$ for the models with
$c_s^2=1$ and (a) $w=-1.2$, (b) $w=-1$, (c) $w=-0.8$, (d) $w=-0.6$, 
(e) $w=-0.4$, respectively.
The solid lines correspond to the numerically integrated solutions, 
whereas the bald dashed lines are derived from the analytic
estimation (\ref{fsigma8}) with the 10-th order terms
of $c_n$. We also show the current RSD data.}
\end{figure}

Let us first study constant $w$ models realized by 
either imperfect fluids or quintessence.
Unless $c_s^2 \ll 1$ the dark energy perturbation is negligible
relative to the matter perturbation, so that the approximate 
equation (\ref{delmeq}) is sufficiently accurate.
Since $w_n=0$ ($n \geq 1$), the coefficients $c_n$ and
$\Omega_{x}$ are given, respectively, by 
\begin{equation}
c_n=\alpha_n=\frac{(-1)^n}{n!}
(\gamma-1)(\gamma-2) \cdots (\gamma-n)\,,\qquad
\Omega_{x}=\frac{\Omega_{x 0} (1+z)^{3w_0}}
{1-\Omega_{x 0}+\Omega_{x 0} (1+z)^{3w_0}}\,.
\label{cnconst}
\end{equation}
For the growth index (\ref{gammaap}) 
we take into account the terms up to 2-nd order with respect to 
$\Omega_{x}$, i.e., 
\begin{equation}
\gamma= \frac{3(1-w_0)}{5-6w_0}+\frac32 
\frac{(1-w_0)(2-3w_0)}{(5-6w_0)^2 (5-12w_0)}
\Omega_{x}+\frac{(w_0-1)(3w_0-2)(324w_0^2-420w_0+97)}
{4(5-6w_0)^3 (5-12w_0)(5-18w_0)}
\Omega_x^2\,.
\end{equation}
Recall that the terms higher than 2-nd order in $\gamma$ 
are negligibly small.
Then $f\sigma_8$ is analytically known from 
Eq.~(\ref{fsigma8}) just by specifying the values of
$\sigma_8 (z=0)$, $\Omega_{x 0}$, and $w=w_0$.

In Fig.~\ref{worder} we plot the evolution of $f\sigma_8$
obtained by the analytic estimation (\ref{fsigma8}) 
for $w=-0.9$ and $c_s^2=1$.
A number of different lines correspond to the solutions 
derived by taking into account the $c_n$ terms up to 
1-st, 3-rd, and 7-th orders.
As we pick up higher-order coefficients $c_n$ in Eq.~(\ref{cnconst}), 
the solutions tend to approach the numerically integrated 
solution of $f \sigma_8$.
In Fig.~\ref{worder} we find that the solution up to 7-th 
order terms of $c_n$ can reproduce
the full numerical result in good precision.

In Fig.~\ref{fsigmafig} we show $f\sigma_8$ versus $z$
for five different values of $w$.
In order to obtain a good convergence
we pick up the $c_n$ terms up to 10-th order.
Figure \ref{fsigmafig} shows that our analytic 
estimation (\ref{fsigma8})
is sufficiently trustable to reproduce the numerically
integrated solutions accurately.
If we only pick up the terms inside $\gamma$ up to 
1-st order with respect to $\Omega_x$, there is small difference
of $f\sigma_8$ between the analytic estimation 
and the numerical solutions for $w \gtrsim -0.6$
(which occurs in the low redshift regime).
Taking into account the 2-nd order term in 
Eq.~(\ref{gammaap}), this difference gets smaller. 
Fig.~\ref{fsigmafig} also displays the RSD data given  
in Table \ref{table:fs8constraints}, which will be used to 
place observational constraints on $w$ in Sec.~\ref{obserconsec}.

\subsection{Tracking quintessence models}
\label{trasec}

From Eq.~(\ref{wtrack2}), we see that all coefficients 
$w_n$'s ($n \geq 1$) in tracking models of quintessence
are expressed in terms of the tracker equation of state 
$w_0=w_{(0)}$. In this case there are contributions to $c_n$ 
coming from the variation of $w$, i.e., non-zero values 
of $\beta_n$. Note that $\beta_n$'s depend only on 
$w_{(0)}$.

\begin{figure}
\includegraphics[height=3.2in,width=3.4in]{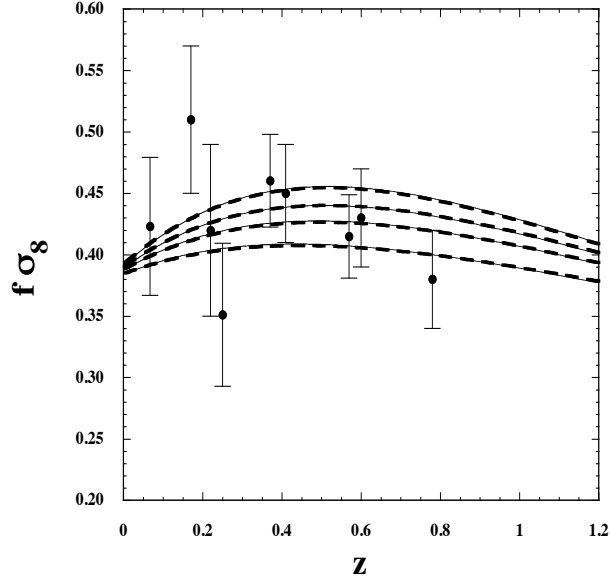}
\caption{\label{tracker}
Evolution of $f\sigma_8$ versus $z$ for the tracking quintessence 
model characterized by the inverse power-law potential
$V(\phi)=M^{4+p}\phi^{-p}$. 
{}From top to bottom the solid lines correspond to the numerically integrated 
solutions for $p=0.1, 0.5, 1, 2$, respectively, whereas the bold dashed lines 
are derived from the analytic solution (\ref{fsigma8}) with the 10-th order terms.
The RSD data are the same as those given in Fig.~\ref{fsigmafig}.}
\end{figure}

For the tracking quintessence with the inverse power-law potential 
$V(\phi)=M^{4+p}\phi^{-p}$ ($p>0$), 
we compare the numerically integrated solutions 
of $f \sigma_8$ with those derived by the analytic expression (\ref{fsigma8}).
In Fig.~\ref{tracker} we show the evolution of $f\sigma_8$ 
for $p=0.1, 0.5, 1, 2$ evaluated from (\ref{fsigma8})
as well as the numerical solutions.
In Eq.~(\ref{fsigma8}) we take into account the $c_n$'s
up to 10-th order terms, whereas in the analytic expression 
of $\gamma$ in Eq.~(\ref{gammaap}) the terms up to 
2-nd order in $\Omega_{x}$ are included.
For the evaluation of $\Omega_{x}$ the 1-st order solution 
(\ref{Omegax1}) with $\Omega_{x0}=0.73$ is used.
{}From Fig.~\ref{tracker} we find that the analytic solution (\ref{fsigma8})
is accurate enough to reproduce the full numerical solution 
in high precisions.
If we take into account the $c_n$'s up to the 3-rd order 
terms, for example, there is some difference between the analytic 
and numerical results. This difference tends to disappear by including 
the higher-order terms of $c_n$.
While the terms up to 10-th order are taken into account 
in Fig.~\ref{tracker}, the 7-th order solutions are 
sufficiently accurate.

While our analytic formula of $f \sigma_8$ is trustable, readers
may think that 7-th order expansion of $c_n$ is not very convenient
for practical purpose. 
However, using this analytic formula is simpler than solving the 
perturbation equations numerically for arbitrary initial conditions.
If we take the latter approach, we need first to identify the present 
epoch (say, $0.7<\Omega_{x0}<0.73$) by solving the background 
equations from some redshift ($z=z_i$). 
Then the perturbation equations are solved with arbitrary initial values 
of $\sigma_8 (z_i)$ to find $f\sigma(z)$ for each $z$.
On the other hand, with our analytic formula, the likelihood analysis 
in terms of 3 free parameters $w_0$, $\sigma_8(z=0)$, and $\Omega_{x0}$
can be done much easier even with the 7-th order expansion of $c_n$.
We also would like to stress that our formula of $f\sigma_8 (z)$
includes the free parameters $\sigma_8 (z=0)$ and  $\Omega_{x0}$ {\it today}, 
by which the joint analysis with other data 
(such as CMB) can be conveniently performed.

\subsection{Thawing models}
\begin{figure}
\includegraphics[height=3.2in,width=3.4in]{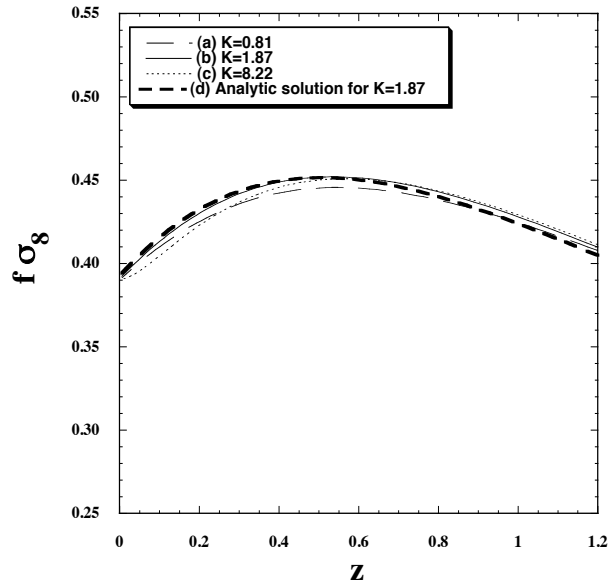}
\caption{\label{thawing}
Evolution of $f\sigma_8$ versus $z$ for the freezing quintessence 
with the potential $V(\phi)=\Lambda^4 [1+\cos (\phi/f)]$.
The cases (a)-(c) correspond to the full numerical solutions
with the model parameters (a) $K=0.81$, $w_1=0.17$, 
(b) $K=1.87$, $w_1=4.7 \times 10^{-2}$, and 
(c) $K=8.22$, $w_1=3.6 \times 10^{-6}$, 
respectively, whereas the case (d) is derived by the 10-th order analytic
solution (\ref{fsigma8}) for $K=1.87$, $w_1=4.7 \times 10^{-2}$.}
\end{figure}

In thawing models of quintessence the field equation of state is given 
by Eq.~(\ref{wtha}). For larger $K$ the deviation of $w$ from $-1$
occurs at later times with a sharper transition.
{}From Eq.~(\ref{wthaw2}) we find that the higher-order terms 
in $\Omega_x$ are not negligible for $K$ larger than the order of 1.
In fact we have numerically confirmed that, for $K>{\cal O}(1)$, 
the expansion (\ref{wthaw2}) does not accommodate
the rapid transition of $w$ at late times unless the higher-order
terms are fully taken into account.
This property also holds for the evolution of $f \sigma_8$.  
Only when $K$ is smaller than the order of 1, the analytic estimation
(\ref{fsigma8}) can reproduce the full numerical solutions
in good precision.

In Fig.~\ref{thawing} we plot the numerical evolution of $f\sigma_8$
for the potential $V(\phi)=\Lambda^4 [1+\cos (\phi/f)]$
with three different values of $K$.
Since $w$ is close to $-1$ in all these cases and the 
deviation from $w=-1$ occurs only at late times, the evolution 
of $f \sigma_8$ is not very different from each other for $K<10$. 
From this analysis, it is clear that, 
only when very accurate data of $f \sigma_8$ are available
in the future, it will be possible to distinguish between the 
models with different values of $K$.
In Fig.~\ref{thawing} we also show the analytic solution derived
for $K=1.87$ and $w_1=4.7 \times 10^{-2}$ as the bold dashed 
line (d). We take into account the $c_n$'s up to
10-th order to evaluate $f\sigma_8$ in Eq.~(\ref{fsigma8}).
Compared to the full numerical solution labelled as (b), 
there is a small difference in the high-redshift regime.
We confirm that this deviation tends to be smaller by involving 
the $c_n$'s higher than 10-th order.
For $K<1$ the analytic estimation is more accurate even without 
including such higher-order terms.

\section{Constraints from the current RSD data}
\label{obserconsec}

In this section, we place observational bounds on two models of dark energy 
discussed in Secs.~\ref{conwsec} and \ref{trasec} by using the current 
RSD data presented in Table \ref{table:fs8constraints}.
For the today's value of $\sigma_8$ we consider 
the prior obtained from observations of CMB, BAO, and Hubble constant 
measurement ($H_0$), i.e.,
\be
\sigma_8 (z=0)=0.816 \pm 0.024\,.
\label{sigprior}
\ee
Here and in what follows all the error bars correspond 
to the 68.3\% confidence level (CL).
Recall that we derived the analytic formula (\ref{fsigma8}) 
under the approximation that the dark energy perturbation 
is neglected relative to the matter perturbation.
For the validity of this approximation we put the prior $w<-0.1$.

\subsection{Constant $w$ models}

For the models of constant $w$ the today's matter density parameter 
constrained from SN Ia, CMB, BAO, and $H_0$ 
observations is \cite{Suzuki}
\be
\Omega_{m0}=0.272^{+0.013}_{-0.013}\,.
\label{Omeprior}
\ee
Under the priors (\ref{sigprior}) and (\ref{Omeprior}) 
we estimate the best-fit to the set of parameters 
${\mathbf{P}} \equiv 
(w, \sigma_8 (z=0), \Omega_{m0})$ by evaluating the likelihood distribution function,
 ${\cal{L}} \propto e^{-\chi^2/2}$, with
\be
\chi^2=\sum_{i=1}^{9} \left( \frac{f \sigma_{8,{\rm ob}}(z_i)
-f \sigma_{8,{\rm th}}(z_i)}
{\sigma_i} \right)^2\; .
\ee
Here $f \sigma_{8,{\rm ob}}(z_i)$ ($i=1,\cdots,9$) are the 9 data displayed 
in Table \ref{table:fs8constraints} with the 
error bars $\sigma_i$, whereas $f \sigma_{8,{\rm th}}(z_i)$
are the theoretical values derived from the analytic solution (\ref{fsigma8}).
For the evaluation of $f \sigma_{8,{\rm th}}(z_i)$ we pick up 
the $c_n$'s up to 10-th order.
For the growth index $\gamma$ the terms up to 
2-nd order with respect to $\Omega_{x}$ are included in Eq.~(\ref{gammaap}).
For $\Omega_x$ we use the 1-st order solution (\ref{Omegax1}).
In tracking quintessence models analyzed in Sec.~\ref{trasubsec} 
we also take the same orders of expansions for $f\sigma_8$, 
$\gamma$, and $\Omega_x$.

We find that the best-fit model parameters are
\be
w=-0.604\,,\qquad \sigma_8 (z=0)=0.840\,,\qquad
\Omega_{m0}=0.285\,,
\ee
with reduced $\chi^2_r=0.947$ ($\chi^2_r \equiv \chi^2_{\rm min}/\nu$, 
where $\nu$ stands for the degrees of freedom).
At 68.3\% CL, our analysis restricts the equation-of-state parameter to the interval
\be
-1.245 < w <-0.347\, ,
\label{bouw}
\ee
whereas $\sigma_8 (z=0)$ and $\Omega_{m0}$ are unconstrained by 
current data even assuming the priors (\ref{sigprior}) and (\ref{Omeprior}).
Although the current bounds on $w$ are weaker than those arising from background tests 
(see, e.g., Refs.~\cite{Komatsu,Suzuki}), 
we expect that upcoming RSD data from ongoing and planned galaxy redshift surveys
can improve this situation in the near future.

\subsection{Tracking quintessence models}
\label{trasubsec}

For the tracking quintessence models in which the equation of state 
is given by Eq.~(\ref{wtrack}) we also carry out the likelihood analysis
by using the analytic solution (\ref{fsigma8}) of $f \sigma_8$ as well as
the expressions for $\gamma$ and $\Omega_x$ 
given in Eqs.~(\ref{gammaap}) and~(\ref{Omegax1}) respectively. 
While the equation of state (\ref{wtrack}) is derived for quintessence, 
we do not impose the prior $w_{(0)}>-1$ for generality. 
For this kind of models, a joint analysis involving current SN Ia, CMB, and BAO gives
the following bound on the matter density parameter \cite{CDT}:
\be
0.273<\Omega_{m0}<0.293\,,
\label{bouome2}
\ee
which is used in our analysis as a prior for $\Omega_{m0}$.
The best-fit model parameters are found to be 
\be
w_{(0)}=-0.461\,,\qquad \sigma_8 (z=0)=0.840\,,\qquad
\Omega_{m0}=0.293\,,
\ee
with $\chi^2_r=0.923$.
At 68.3\% CL, we found
\be
-1.288 < w_{(0)} <-0.214\,,
\label{bouw2}
\ee
whereas the parameters $\sigma_8 (z=0)$ and $\Omega_{m0}$ 
are again unconstrained in the regions of 
(\ref{sigprior}) and (\ref{bouome2}). 
As expected, the bounds on $w_{(0)}$ are weaker than those obtained 
in constant $w$ models [Eq.~(\ref{bouw})]. 
We note that in tracking models the equation of state decreases 
at late times, which is accompanied by the decrease 
of $f\sigma_8$.
Compared to constant $w$ models, this
allows the possibility to fit the data better
even for larger values of $w$ during the matter era.

\section{Conclusions}
\label{consec}

In this paper we have provided an analytic formula of $f \sigma_8$ 
for dynamical dark energy models in the framework of GR.
This was derived by using the approximate matter 
perturbation equation (\ref{delmeq}), which is trustable 
as long as the contribution of the dark energy perturbation to 
the gravitational potential is negligible relative to that 
of the matter perturbation. 
Our formula of $f \sigma_8$ can be applied to many dark 
energy models including imperfect fluids, quintessence, 
and k-essence in which the sound speed squared $c_s^2$
is not very close to 0.

Our derivation of $f \sigma_8$ is based on the expansion 
of $w$ with respect to the dark energy density parameter 
$\Omega_x$, i.e., $w=w_0+\sum_{n=1} w_n (\Omega_x)^n$.
The growth rate $f=\delta_m'/\delta_m$ of 
matter perturbations is parametrized by the growth index 
$\gamma$, as $f=(1-\Omega_x)^{\gamma}$.
We expanded $\gamma$ in terms of $\Omega_x$ up to 
2-nd order terms.
Since $\gamma$ is dominated by the term $\gamma_0=3(1-w_0)/(5-6w_0)$, 
it is a good approximation to treat $\gamma$ as a constant for 
the derivation of the integrated solution of $\delta_m$.
The $c_n$'s in Eq.~(\ref{fsigma8}) are given by Eq.~(\ref{cndef}), 
where $\alpha_n$ and $\beta_n$ appear as the coefficients 
of the expansion of the terms $(1-\Omega_x)^{\gamma-1}$
and $1/w$ respectively.
For the density parameter $\Omega_x$, the 1-st order solution 
(\ref{Omegax1}) is usually sufficient to obtain
accurate analytic solutions of $f\sigma_8$.

In Sec.~\ref{validitysec} we have studied the validity of the analytic formula
(\ref{fsigma8}) in concrete models of dark energy.
For constant $w$ models in which $c_n$ and $\Omega_x$ are
given by Eq.~(\ref{cnconst}), the analytic solution up to 7-th order
terms of $c_n$ reproduces the numerically integrated solutions with good precision.
This property also holds for tracking quintessence models where 
the evolution of $w$ is given by Eq.~(\ref{wtrack2}).
In thawing quintessence and k-essence models, where $w$
is given by Eq.~(\ref{wtha}), the formula (\ref{fsigma8}) 
can be trustable for $K \lesssim 1$, but for $K$ larger than 
the order of 1, we need to fully take into account the higher-order 
terms of $c_n$ to have good convergence of $f \sigma_8$.
 
In Sec.~\ref{obserconsec} we have discussed observational constraints on 
two different dark energy models by using the current RSD data.
In both constant $w$ and tracking quintessence models the analytic 
solution (\ref{fsigma8}) includes the three parameters $\sigma_8(z=0)$, 
$\Omega_{m0}$, and $w$ (or $w_{(0)}$).
Under the priors on $\sigma_8(z=0)$ and $\Omega_{m0}$ constrained
by SN Ia, CMB, BAO, and $H_0$ measurements, we derived the 
bounds $-1.245<w<-0.347$ (68\,\%\,CL) for constant $w$ models and 
$-1.288<w_{(0)}<-0.214$ (68\,\%\,CL) for tracking quintessence models.
Although the upper bounds on the dark energy equation of state 
are still weak with current data, we expect to obtain 
more precise data from ongoing surveys or near-future projects
such as Subaru/FMOS, HETDEX, and J-PAS.
Our analytic formula of $f \sigma_8$ will be useful to place tighter 
bounds on dynamical dark energy models in the future.

So far, observational bounds on $f\sigma_8$ 
(listed in Table~\ref{table:fs8constraints}) have been derived in the 
standard cosmological scenario without taking into 
account additional effects such as a possible coupling between dark 
matter and dark energy, irrotational flow, and so on.
Reflecting this observational status, we did not assume any non-standard 
picture to estimate the theoretical values of $f \sigma_8$.
However, if the standard cosmological scenario does not match with 
future high-precision data very well, it may be necessary to
include non-standard effects mentioned above as a next step.
We leave the theoretical estimation of such effects for future work.

\section*{ACKNOWLEDGEMENTS}
We thank Luca Amendola, Hiroyuki Okada, and 
Tomonori Totani for useful discussions.
A.~D.~F.\ is supported by JSPS (under the grant No.~S12135)
and thanks Tokyo University of Science, for the warm hospitality 
received while part of the project was finalized.
S.~T.\ is supported by the Grant-in-Aid for
Scientific Research Fund of the Fund of the 
JSPS No.~24540286 and Scientific Research 
on Innovative Areas (No.~21111006). 
S.~T.\ thanks warm hospitalities during his stays in 
Weihai, Observatorio Nacional in Rio de Janeiro, 
Passa Quatro, Szczecin, and University of Heidelberg. 
J.~S.~A. is supported by CNPq under Grants No.~305857/2010-0 
and No.~485669/2011-0 and FAPERJ Grant No.~E-26/103.239/2011.

\appendix


\begin{thebibliography}{10}

\bibitem{RP}
A.~G.~Riess {\it et al.}  [Supernova Search Team Collaboration],
Astron.\ J.\  {\bf 116}, 1009 (1998);
S.~Perlmutter {\it et al.}  [Supernova Cosmology Project Collaboration],
Astrophys.\ J.\  {\bf 517}, 565 (1999).

\bibitem{Spergel} 
D.~N.~Spergel {\it et al.}  [WMAP Collaboration],
Astrophys.\ J.\ Suppl.\  {\bf 148}, 175 (2003).

\bibitem{Komatsu} 
E.~Komatsu {\it et al.}  [WMAP Collaboration],
Astrophys.\ J.\ Suppl.\  {\bf 192}, 18 (2011).

\bibitem{BAO}
D.~J.~Eisenstein {\it et al.}  [SDSS Collaboration],
of SDSS luminous red galaxies,''
Astrophys.\ J.\  {\bf 633}, 560 (2005);
W.~J.~Percival {\it et al.}  [SDSS Collaboration],
Mon.\ Not.\ Roy.\ Astron.\ Soc.\  {\bf 401}, 2148 (2010).

\bibitem{Weinberg}
S.~Weinberg,
Rev.\ Mod.\ Phys.\  {\bf 61}, 1 (1989).

\bibitem{review}
V.~Sahni and A.~A.~Starobinsky,
Int.\ J.\ Mod.\ Phys.\ D \textbf{9}, 373 (2000).

\bibitem{review2}
S.~M.~Carroll,
Living Rev.\ Rel.\  {\bf 4}, 1 (2001);
T.~Padmanabhan,
Phys.\ Rept.\ \textbf{380}, 235 (2003);
P.~J.~E.~Peebles and B.~Ratra,
Rev.\ Mod.\ Phys.\  {\bf 75}, 559 (2003);
E.~J.~Copeland, M.~Sami and S.~Tsujikawa,
Int.\ J.\ Mod.\ Phys.\ D {\bf 15}, 1753 (2006);
T.~P.~Sotiriou and V.~Faraoni,
Rev.\ Mod.\ Phys.\  {\bf 82}, 451 (2010);
A.~De Felice and S.~Tsujikawa,
Living Rev.\ Rel.\  {\bf 13}, 3 (2010);
S.~Tsujikawa,
Lect.\ Notes Phys.\  {\bf 800}, 99 (2010);
arXiv:1004.1493 [astro-ph.CO].

\bibitem{Suzuki}
N.~Suzuki  {\it et al.},
Astrophys.\ J.\  {\bf 746}, 85 (2012).

\bibitem{BlakeBAO} 
C.~Blake {\it et al.},
Mon.\ Not.\ Roy.\ Astron.\ Soc.\  {\bf 418}, 1707 (2011).

\bibitem{BOSS12} 
L.~Anderson  {\it et al.},
arXiv:1203.6594 [astro-ph.CO].

\bibitem{DNT12} 
A.~De Felice, S.~Nesseris and S.~Tsujikawa,
JCAP {\bf 1205}, 029 (2012).

\bibitem{quinpapers}
Y.~Fujii, Phys.\ Rev.\ D {\bf 26}, 2580 (1982);
L.~H.~Ford,
Phys.\ Rev.\ D {\bf 35}, 2339 (1987);
C.~Wetterich, Nucl. \ Phys \ B. {\bf 302},
668 (1988);
T.~Chiba, N.~Sugiyama and T.~Nakamura,
Mon.\ Not.\ Roy.\ Astron.\ Soc.\  {\bf 289}, L5 (1997);
P.~G.~Ferreira and M.~Joyce,
Phys.\ Rev.\ Lett.\  {\bf 79}, 4740 (1997);
R.~R.~Caldwell, R.~Dave and P.~J.~Steinhardt,
Phys.\ Rev.\ Lett.\  {\bf 80}, 1582 (1998).

\bibitem{kinf} 
C.~Armendariz-Picon, T.~Damour and V.~F.~Mukhanov,
Phys.\ Lett.\ B {\bf 458}, 209 (1999).

\bibitem{kespapers} 
T.~Chiba, T.~Okabe and M.~Yamaguchi, 
Phys.\ Rev.\ {\bf D62}, 023511 (2000); 
C.~Armendariz-Picon,
V.~F.~Mukhanov and P.~J.~Steinhardt, 
Phys.\ Rev.\ Lett.\ {\bf 85}, 4438-4441 (2000).

\bibitem{fR} 
S.~Capozziello, 
Int.\ J.\ Mod.\ Phys.\ D {\bf 11}, 483 (2002); 
S.~Capozziello, S.~Carloni and A.~Troisi, 
Recent Res.\ Dev.\ Astron.\ Astrophys.\ {\bf 1}, 625 (2003);
S.~M.~Carroll, V.~Duvvuri, M.~Trodden and M.~S.~Turner, 
Phys.\ Rev.\ D {\bf 70}, 043528 (2004);
L.~Amendola, R.~Gannouji, D.~Polarski and S.~Tsujikawa,
Phys.\ Rev.\ D {\bf 75}, 083504 (2007);
L.~Amendola and S.~Tsujikawa,
Phys.\ Lett.\ B {\bf 660}, 125 (2008);
W.~Hu and I.~Sawicki,
Phys.\ Rev.\ D {\bf 76}, 064004 (2007);
A.~A.~Starobinsky,
JETP Lett.\  {\bf 86}, 157 (2007);
S.~A.~Appleby and R.~A.~Battye,
Phys.\ Lett.\ B {\bf 654}, 7 (2007);
S.~Tsujikawa,
Phys.\ Rev.\ D {\bf 77}, 023507 (2008);
E.~V.~Linder,
Phys.\ Rev.\ D {\bf 80}, 123528 (2009); 
M.~Campista, B.~Santos, J.~Santos and J.~S.~Alcaniz,
Phys.\ Lett.\ B {\bf 699}, 320 (2011);
B.~Santos, M.~Campista, J.~Santos and J.~S.~Alcaniz,
arXiv:1207.2478 [astro-ph.CO].

\bibitem{Tegmark03} 
M.~Tegmark {\it et al.}  [SDSS Collaboration],
Phys.\ Rev.\ D {\bf 69}, 103501 (2004);
U.~Seljak {\it et al.}  [SDSS Collaboration],
Phys.\ Rev.\ D {\bf 71}, 103515 (2005);
M.~Tegmark {\it et al.}  [SDSS Collaboration],
Phys.\ Rev.\ D {\bf 74}, 123507 (2006).

\bibitem{Kaiser} 
N.~Kaiser,
Mon.\ Not.\ Roy.\ Astron.\ Soc.\  {\bf 227}, 1 (1987).

\bibitem{Tegmark04} 
M.~Tegmark {\it et al.}  [SDSS Collaboration],
Astrophys.\ J.\  {\bf 606}, 702 (2004).

\bibitem{Percival04}
W.~J.~Percival {\it et al.}  [The 2dFGRS Collaboration],
Mon.\ Not.\ Roy.\ Astron.\ Soc.\  {\bf 353}, 1201 (2004).

\bibitem{Porto} 
C.~Di Porto and L.~Amendola,
Phys.\ Rev.\ D {\bf 77}, 083508 (2008).

\bibitem{Guzzo08} 
L.~Guzzo  {\it et al.},
Nature {\bf 451}, 541 (2008).

\bibitem{Savas} 
S.~Nesseris and L.~Perivolaropoulos,
Phys.\ Rev.\ D {\bf 77}, 023504 (2008).

\bibitem{Alam} 
U.~Alam, V.~Sahni and A.~A.~Starobinsky,
Astrophys.\ J.\  {\bf 704}, 1086 (2009).

\bibitem{Blake} 
C.~Blake {\it et al.},
Mon.\ Not.\ Roy.\ Astron.\ Soc.\  {\bf 415}, 2876 (2011).

\bibitem{Samushia11} 
L.~Samushia, W.~J.~Percival and A.~Raccanelli,
Mon.\ Not.\ Roy.\ Astron.\ Soc.\  {\bf 420}, 2102 (2012).

\bibitem{Reid12} 
B.~A.~Reid {\it et al.},
arXiv:1203.6641 [astro-ph.CO].

\bibitem{Beutler12} 
F.~Beutler {\it et al.},
arXiv:1204.4725 [astro-ph.CO].

\bibitem{Koyama} 
E.~Jennings, C.~M.~Baugh, B.~Li, G.~-B.~Zhao and K.~Koyama,
arXiv:1205.2698 [astro-ph.CO].

\bibitem{Rapetti} 
D.~Rapetti, C.~Blake, S.~W.~Allen, A.~Mantz, D.~Parkinson and F.~Beutler,
arXiv:1205.4679 [astro-ph.CO].

\bibitem{Samushia12} 
L.~Samushia {\it et al.},
arXiv:1206.5309 [astro-ph.CO].

\bibitem{Linder12} 
S.~A.~Appleby and E.~V.~Linder,
JCAP {\bf 1208}, 026 (2012).

\bibitem{OTT12} 
H.~Okada, T.~Totani and S.~Tsujikawa,
arXiv:1208.4681 [astro-ph.CO].

\bibitem{fRper}
S.~M.~Carroll, I.~Sawicki, A.~Silvestri and M.~Trodden,
New J.\ Phys.\  \textbf{8}, 323 (2006);
R.~Bean, D.~Bernat, L.~Pogosian, A.~Silvestri and M.~Trodden,
Phys.\ Rev.\  D \textbf{75}, 064020 (2007);
S.~Tsujikawa,
Phys.\ Rev.\ D {\bf 76}, 023514 (2007);
L.~Pogosian and A.~Silvestri,
Phys.\ Rev.\ D {\bf 77}, 023503 (2008);
S.~Tsujikawa, R.~Gannouji, B.~Moraes and D.~Polarski,
Phys.\ Rev.\ D {\bf 80}, 084044 (2009).

\bibitem{Kase}
A.~De Felice, R.~Kase and S.~Tsujikawa,
Phys.\ Rev.\ D {\bf 83}, 043515 (2011).

\bibitem{DTextended}
A.~De Felice and S.~Tsujikawa,
JCAP {\bf 1203}, 025 (2012).

\bibitem{Zlatev} 
I.~Zlatev, L.~M.~Wang and P.~J.~Steinhardt,
Phys.\ Rev.\ Lett.\  {\bf 82}, 896 (1999);
P.~J.~Steinhardt, L.~-M.~Wang and I.~Zlatev,
Phys.\ Rev.\ D {\bf 59}, 123504 (1999).

\bibitem{Bardeen} 
J.~M.~Bardeen,
Phys.\ Rev.\ D {\bf 22}, 1882 (1980).

\bibitem{Kodama} 
H.~Kodama and M.~Sasaki,
Prog.\ Theor.\ Phys.\ Suppl.\  {\bf 78}, 1 (1984).

\bibitem{Bean} 
R.~Bean and O.~Dore,
Phys.\ Rev.\ D {\bf 69}, 083503 (2004).

\bibitem{Sapone}
D.~Sapone, M.~Kunz and L.~Amendola,
Phys.\ Rev.\ D {\bf 82}, 103535 (2010).

\bibitem{DEbook}
L.~Amendola and S.~Tsujikawa, 
{\it Dark energy--Theory and Observations}, 
Cambridge University Press (2010).

\bibitem{BTW05}
B.~A.~Bassett, S.~Tsujikawa and D.~Wands,
Rev.\ Mod.\ Phys.\  {\bf 78}, 537 (2006).

\bibitem{Garriga} 
J.~Garriga and V.~F.~Mukhanov,
Phys.\ Lett.\ B {\bf 458}, 219 (1999).

\bibitem{Tegmark06} 
M.~Tegmark {\it et al.}  [SDSS Collaboration],
Phys.\ Rev.\ D {\bf 74}, 123507 (2006).

\bibitem{Song09} 
Y.~-S.~Song and W.~J.~Percival,
JCAP {\bf 0910}, 004 (2009).

\bibitem{White} 
M.~White, Y.~-S.~Song and W.~J.~Percival,
Mon.\ Not.\ Roy.\ Astron.\ Soc.\  {\bf 397}, 1348 (2008).

\bibitem{Riotto} 
G.~Ballesteros and A.~Riotto,
Phys.\ Lett.\ B {\bf 668}, 171 (2008).

\bibitem{Saini03} 
V.~Sahni, T.~D.~Saini, A.~A.~Starobinsky and U.~Alam,
JETP Lett.\  {\bf 77}, 201 (2003).

\bibitem{Sahni06} 
V.~Sahni and A.~Starobinsky,
Int.\ J.\ Mod.\ Phys.\ D {\bf 15}, 2105 (2006).

\bibitem{CL05} 
R.~R.~Caldwell and E.~V.~Linder,
Phys.\ Rev.\ Lett.\  {\bf 95}, 141301 (2005).

\bibitem{Ratra}
P.~J.~E.~Peebles and B.~Ratra,
Astrophys.\ J.\  {\bf 325}, L17 (1988);
B.~Ratra and J.~Peebles,
Phys. \ Rev \ D {\bf 37}, 3406 (1988).

\bibitem{Chiba} 
T.~Chiba,
Phys.\ Rev.\ D {\bf 81}, 023515 (2010).

\bibitem{PNGB}
J.~A.~Frieman, C.~T.~Hill, A.~Stebbins and I.~Waga,
Phys.\ Rev.\ Lett.\  {\bf 75}, 2077 (1995).

\bibitem{Dutta} 
S.~Dutta and R.~J.~Scherrer,
Phys.\ Rev.\ D {\bf 78}, 123525 (2008).

\bibitem{Scherrer} 
R.~J.~Scherrer and A.~A.~Sen,
Phys.\ Rev.\ D {\bf 77}, 083515 (2008).

\bibitem{Chibake}
T.~Chiba, S.~Dutta and R.~J.~Scherrer,
Phys.\ Rev.\ D {\bf 80}, 043517 (2009).

\bibitem{Piazza} 
F.~Piazza and S.~Tsujikawa,
JCAP {\bf 0407}, 004 (2004).

\bibitem{Mukohyama} 
N.~Arkani-Hamed, H.~-C.~Cheng, M.~A.~Luty and S.~Mukohyama,
JHEP {\bf 0405}, 074 (2004).

\bibitem{Wang}
L.~-M.~Wang and P.~J.~Steinhardt,
Astrophys.\ J.\  {\bf 508}, 483 (1998).

\bibitem{Peebles}
P.~J.~E.~Peebles, {\it Large-Scale Structure of the Universe,}
Princeton University Press (1980).

\bibitem{Gong} 
Y.~Gong, M.~Ishak and A.~Wang,
Phys.\ Rev.\ D {\bf 80}, 023002 (2009).

\bibitem{Linder} 
E.~V.~Linder,
Phys.\ Rev.\ D {\bf 72}, 043529 (2005);
E.~V.~Linder and R.~N.~Cahn,
Astropart.\ Phys.\  {\bf 28}, 481 (2007).

\bibitem{CDT} 
T.~Chiba, A.~De Felice, and S.~Tsujikawa, 
arXiv:1210.3859  [astro-ph.CO].

\end{thebibliography}
\end{document}